\documentclass[aps,pra,twocolumn,superscriptaddress,showpacs]{revtex4-1}

\usepackage{inputenc}
\usepackage{color}
\usepackage{graphicx}
\usepackage{amsmath}
\usepackage{amssymb}
\usepackage{float}

\newcommand{\dd}{\ensuremath{\mathrm{d}}}

\newcommand{\mean}[1]{\ensuremath{\left\langle #1\right\rangle}}
\newcommand{\pdf}{\ensuremath{\operatorname{PDF}}}
\newcommand{\erf}{\ensuremath{\operatorname{erf}}}

\newcommand{\pdfkde}{\ensuremath{\operatorname{PDF}^\mathtt{KDE}}}
\newcommand{\pdfn}{\ensuremath{\operatorname{PDF}^\mathtt{N}}}
\newcommand{\pdfht}{\ensuremath{\operatorname{PDF}^\mathtt{HT}}}

\def\3He{$^3$He} 
\def\4He{$^4$He}

\definecolor{clr0}{rgb}{0.122,0.467,0.706}
\definecolor{clr1}{rgb}{1.000,0.498,0.055}
\definecolor{clr2}{rgb}{0.173,0.627,0.173}
\definecolor{clr3}{rgb}{0.839,0.153,0.157}
\definecolor{clr4}{rgb}{0.580,0.404,0.741}
\definecolor{clr5}{rgb}{0.549,0.337,0.294}
\definecolor{clr6}{rgb}{0.890,0.467,0.761}
\definecolor{clr7}{rgb}{0.498,0.498,0.498}
\definecolor{clr8}{rgb}{0.737,0.741,0.133}
\definecolor{clr9}{rgb}{0.090,0.745,0.812}

\definecolor{pblue}{rgb}{0.122,0.467,0.706}
\definecolor{porange}{rgb}{1.000,0.498,0.055}
\definecolor{pgreen}{rgb}{0.173,0.627,0.173}
\definecolor{pred}{rgb}{0.839,0.153,0.157}
\definecolor{pviolet}{rgb}{0.580,0.404,0.741}
\definecolor{pbrown}{rgb}{0.549,0.337,0.294}
\definecolor{ppink}{rgb}{0.890,0.467,0.761}
\definecolor{pgrey}{rgb}{0.498,0.498,0.498}
\definecolor{pyellow}{rgb}{0.737,0.741,0.133}
\definecolor{pcyan}{rgb}{0.090,0.745,0.812}

\newcommand{\pcircle}{$\bullet$}
\newcommand{\psquare}{$\blacksquare$}
\newcommand{\ptriup}{$\blacktriangle$}
\newcommand{\ptridown}{$\blacktriangledown$}

\newcommand{\pdiamond}{$\blacklozenge$}

\begin{document}

\title{Intermittency enhancement in quantum turbulence}
\author{Emil~Varga}
\email[Email: ]{varga.emil@gmail.com}
\affiliation{National High Magnetic Field Laboratory, 1800 East Paul Dirac Drive, Tallahassee, Florida 32310, USA}
\affiliation{Faculty of Mathematics and Physics, Charles University, Ke Karlovu 3, 121 16, Prague, Czech Republic}

\author{Jian~Gao}
\affiliation{National High Magnetic Field Laboratory, 1800 East Paul Dirac Drive, Tallahassee, Florida 32310, USA}
\affiliation{Mechanical Engineering Department, Florida State University, Tallahassee, Florida 32310, USA}

\author{Wei~Guo}
\email[Email: ]{wguo@magnet.fsu.edu}
\affiliation{National High Magnetic Field Laboratory, 1800 East Paul Dirac Drive, Tallahassee, Florida 32310, USA}
\affiliation{Mechanical Engineering Department, Florida State University, Tallahassee, Florida 32310, USA}

\author{Ladislav~Skrbek}
\affiliation{Faculty of Mathematics and Physics, Charles University, Ke Karlovu 3, 121 16, Prague, Czech Republic}
\date{\today}

\begin{abstract}
  Intermittency is a hallmark of turbulence, which exists not only in turbulent flows of classical viscous fluids but also in flows of quantum fluids
  such as superfluid $^4$He. Despite the established similarity between turbulence in classical fluids and quasi-classical turbulence in superfluid
  $^4$He, it has been predicted that intermittency in superfluid $^4$He is temperature dependent and enhanced for certain temperatures, which
  strikingly contrasts the nearly flow-independent intermittency in classical turbulence.  Experimental verification of this theoretical prediction is
  challenging since it requires well-controlled generation of quantum turbulence in $^4$He and flow measurement tools with high spatial and temporal
  resolution. Here, we report an experimental study of quantum turbulence generated by towing a grid through a stationary sample of superfluid
  $^4$He. The decaying turbulent quantum flow is probed by combining a recently developed He$^*_2$ molecular tracer-line tagging velocimetry technique
  and a traditional second sound attenuation method. We observe quasi-classical decays of turbulent kinetic energy in the normal fluid and of vortex
  line density in the superfluid component. For several time instants during the decay, we calculate the transverse velocity structure
  functions. Their scaling exponents, deduced using the extended self-similarity hypothesis, display non-monotonic temperature-dependent intermittency
  enhancement, in excellent agreement with recent theoretical/numerical study of Biferale \emph{et al.}  [Phys. Rev. Fluids 3, 024605 (2018)].
\end{abstract}

\maketitle

\section{Introduction}
Intermittency in turbulent flows is a topic of extensive study in classical fluid dynamics research \cite{Kolmogorov-JFM-1962, Meneveau-PRL-1987,
  Sreenivasan-ARFM-1997, Biferale-ARFM-2003, Benzi-JSP-2015}. In fully developed turbulence, intermittency manifests itself as extreme velocity
excursions that appear more frequently than one would expect on the basis of Gaussian statistics. Small-scale intermittency results in corrections to
the energy spectrum and velocity structure functions that are nearly universal across a wide range of turbulent flows in classical fluids
\cite{She-PRL-1994, Schumacher-PNAS-2014}. A question that has attracted increasing interest in recent years is whether this universality can be
extended to quantum fluids such as superfluid $^4$He whose hydrodynamic behavior is strongly affected by quantum effects and cannot be described by
the Navier-Stokes equation \cite{Maurer-EPJ-1998,Salort2011a,Boue2013a,Shukla-PRE-2016,Biferale2017,Rusaouen2017}.

Below about $T_\lambda\simeq 2.17$~K, liquid $^4$He undergoes a second order phase transition into a superfluid phase called He II. According to the
two fluid model \cite{tilley_book}, He II behaves as if it is composed of two interpenetrating liquids -- a superfluid component and a normal-fluid
component made off thermal excitations called phonons and rotons. While the normal fluid behaves classically, possessing finite viscosity and carrying
the entire entropy content of He II, the superfluid component has neither entropy nor viscosity. Due to quantum restriction, vorticity in the
superfluid is constrained into line singularities, each carrying a single quantum of circulation $\kappa \approx 9.97\times 10^{-4}$ cm$^2$/s around
its angstrom-sized core \cite{donnelly_book}. The fraction ratio of the two fluids strongly depends on temperature. Above 1 K where both fluids are
present, turbulence in He II (also termed as quantum turbulence \cite{Barenghi2014b}) takes the form of a tangle of quantized vortices in the
superfluid component, co-existing with more classical-like turbulent flow of the normal fluid. When the velocity fields of the two fluids are
mismatched, a mutual friction force between them, arising from the scattering of thermal excitations off the cores of quantized vortices, provides an
inter-component energy transfer and additional dissipation, resulting in a modified turbulence scaling
\cite{Vinen2002,Skrbek2012,Marakov2014,Khomenko2016,Gao2017}.

The general properties of quantum turbulence in He II above 1 K depend on the type of forcing. When the turbulence is generated by an applied heat
current in He II, the two fluids are forced to move with opposite mean velocities (i.e., thermal counterflow) \cite{tilley_book}. The mutual friction
acts at all length scales in both fluids which leads to strongly non-classical behavior and decay \cite{Gao2016a,Babuin2016,Gao2017}. On the other
hand, when the turbulence is generated by methods conventionally used in classical fluid dynamics research, such as by a towed grid
\cite{Stalp1999,Skrbek2000} or using counter-rotating propellers \cite{Maurer-EPJ-1998}, the two fluids can become strongly coupled by the mutual
friction force at large scales and behave like a single-component fluid (i.e., quasi-classical turbulence), possessing some effective viscosity
\cite{Khomenko2016, Gao-PRB-2016}. This coupling must break down at scales comparable or smaller than the mean inter-vortex distance $\ell_Q=L^{-1/2}$
(where $L$ denotes the vortex line density, i.e., the vortex line length per unit volume) since the flow of the superfluid component at these small
scales is restricted to individual vortex lines and cannot match the velocity field of the normal fluid \cite{Gao-PRB-2018}. The quantity $\ell_Q$ is
also known as the ``quantum length scale''; it scales similarly to the Kolmogorov dissipation scale, $\eta$, of classical turbulence
\cite{Babuin2014a}.

The similarity between quasi-classical turbulence in He II and turbulence in classical fluids has attracted a great deal of interest in both quantum
and classical fluid dynamics research fields \cite{Skrbek-JPCM-1999,Donnelly_book-1991}. Extensive experimental, theoretical, and numerical work has
been conducted to explore various properties of turbulence in He II (see the reviews~\cite{Skrbek2012,Barenghi-PNAS-2014} and references therein). In
recent years, intermittency in He II quasi-classical turbulence has become one of the central topics. Since the coupling of the two fluids at large
scales and their decoupling at small scales are all controlled by the temperature dependent mutual friction, one may naturally expect temperature
dependent turbulence statistics. Indeed, it has been predicted by Bou\'{e} \emph{et al.}  \cite{Boue2013a} and Biferale \emph{et al.}
\cite{Biferale2017} that when probed at small scales, intermittency corrections to the scaling of higher-order velocity structure functions in He II
quasi-classical turbulence should be enhanced in the temperature range $1.3 \lesssim T \lesssim 2.1$ K, with a maximum deviation from the
Kolmogorov-Obukhov K41 theory for classical turbulence \cite{Kolmogorov1941} around 1.85~K. Early experiments conducted at low temperatures and close
to $T_\lambda$ did not find deviations from the statistics of classical turbulence \cite{Roche-EPL-2009,Salort-PF-2010,Salort2011a}. A more recent
experiment in a turbulent wake in He II covered a wider range of temperatures but also reported temperature independent intermittency, similar to that
in classical flows \cite{Rusaouen2017}. It should be noted, however, that the pressure and velocity probes used in these experiments all have sizes
much larger than $\ell_Q$ and hence are sensitive only for the corresponding part of the turbulent cascade \cite{Rusaouen2017, Biferale2017}.

A reliable determination of intermittency in He II requires not only the generation of fully developed turbulence but also flow measurement tools with
a spatial resolution comparable to $\ell_Q$. In this paper, we report an experimental study of quasi-classical turbulence generated by towing a grid
through a stationary sample of He II. The velocity of the normal fluid is measured using a recently developed He$^*_2$ molecular tracer-line tagging
velocimetry technique \cite{Marakov2014,Gao2015} while the vortex line density in the superfluid component is determined using a traditional second
sound attenuation method \cite{Stalp1999, Vinen-PRC-1957}. Our experimental results indeed demonstrate intermittency enhancement, in excellent
agreement with the theory predictions \cite{Boue2013a,Biferale2017}.

\section{Experimental Method}
\begin{figure}
  \centering
  \includegraphics[width=0.9\linewidth]{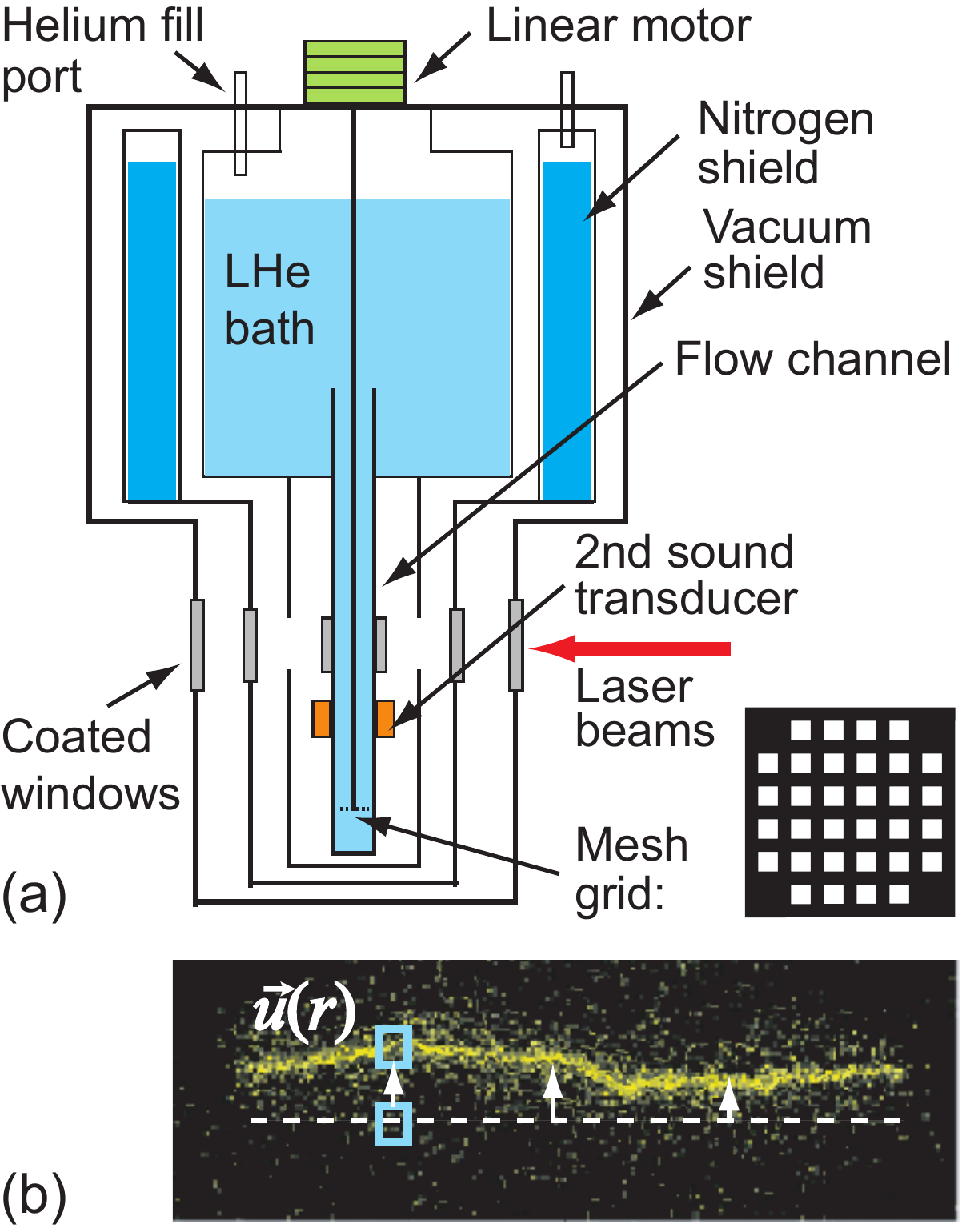}%
  \caption{(a) Schematic diagram of the experimental setup. (b) A sample image of the He$^*_2$ molecular tracer line. The white dashed line serves to
    demonstrate the initial location of the trace line for velocity calculations.}
  \label{Fig1}
\end{figure}

The experiment utilizes the Tallahassee He$_2^*$ tracer-line visualization setup \cite{Gao2015} as shown schematically in Fig. \ref{Fig1} (a). A
stainless steel channel (inner cross-section: 9.5$\times$9.5 mm$^2$; length 300 mm) is attached to a pumped helium bath whose temperature can be
controlled within 0.1 mK. A mesh grid of $7 \times 7$ woven wires (about 8 mm in length and 0.41 mm in thickness) is supported inside the channel at
the four corners and can be towed by a linear motor to move past our flow probes at a controlled speed up to about 65 cm/s. The grid is designed to
have an open area of 54\% so as to avoid producing secondary flows \cite{Fernando-PFA-1993}. The flow generated in the wake of a moving grid is
usually treated as a prototype of nearly homogeneous and isotropic turbulence, the simplest form of turbulence that has been extensively studied in
classical fluid dynamics research \cite{ComteBellot1966,Tennekes1972book,Skrbek2000,Sinhuber2015}. The grid turbulence has also been utilized as a
valuable vantage point in quantum turbulence research for assessing the similarities and differences between classical and quantum turbulent flows
\cite{Stalp1999,Babuin2014a, Stalp-PF-2002}.

To probe the flow, we send high-intensity femtosecond laser pulses through the channel via a pair of slits (about 1 mm in width and 10 mm in length)
cut into opposite sides of the channel along its length. These slits are covered with indium sealed extension flanges and windows. As a consequence of
femtosecond laser-field ionization \cite{Benderskii1999}, a thin line of He$_2^*$ molecular tracers can be created along the beam path
\cite{Gao2015}. The initial thickness of the He$_2^*$ tracer line is about 100 $\mu$m and its length matches the channel width. Above about 1~K, these
He$_2^*$ molecular tracers are completely entrained by the viscous normal fluid with negligible effect from the superfluid or quantized vortices
\cite{Zmeev2013}. A line of the molecules so created is then left to evolve for a drift time $t_{d}$ of about 10--30~ms before it is visualized by
laser-induced fluorescence using a separate laser sheet at 905 nm for imaging \cite{Gao2015}. The streamwise velocity $v_y(x)$ can be determined by
dividing the displacement of a line segment at $x$ by $t_{d}$ (see Fig. \ref{Fig1} (b)). The transverse velocity increments
$\delta v_y(r)=v_y(x)-v_y(x+r)$ can thus be evaluated for structure function calculations. Additionally, the flow is also probed by a standard
second-sound attenuation method \cite{Babuin2012,Gao2015}, revealing temporal decay of vortex line density $L(t)$ in the superfluid.

The grid starts moving from about 50 mm below the second sound sensors up to the uppermost position which is roughly 100 mm above the 1 cm $\times$1
cm visualisation region. Since no steady input of energy into the flow exists (except marginal parasitic radiative heat leaks), the flow starts to
decay after the passage of the grid. As the origin of time for both visualization and second sound data, we take the instant when the grid passes the
position where a tracer line would be inscribed. To study the time evolution, tracer line inscription is delayed until the desired decay time $t$. The
measurement at each decay time is normally repeated $100-200$ times for statistical analysis, and every time the grid is towed anew. The experiments
were performed in a temperature range $1.45-2.15$ K with quadratically increasing decay times (typically) 1, 2, 4, and 8 s. In all cases, the grid
velocity $v_g$ was set to either 300 or 50 mm/s.

\section{Experimental Results}
\subsection{Temporal evolution of the grid turbulence}
\begin{figure}
  \centering
  \includegraphics{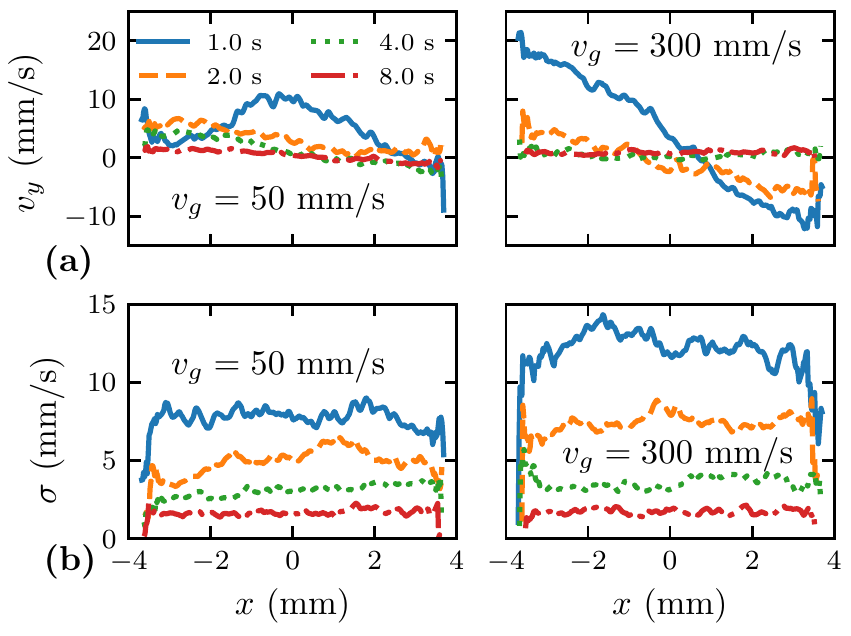}%
  \caption{(a) The ensemble-averaged velocity profile $v_y(x)$ across the channel at different decay times with grid velocities $v_g$ as
    indicated. (b) The corresponding velocity variance $\sigma(x)$ profiles. The shown data are obtained at 1.85 K, at the indicated time instants.}
  \label{fig:tracer-lines}
\end{figure}

In Fig~\ref{fig:tracer-lines}, we show the profiles of the mean velocity $\overline{v_y}(x)=\langle{v_y(x)}\rangle_x$ and the velocity variance
$\sigma(x)=\langle\left(v_y(x)-\overline{v_y}\right)^2\rangle_x^{1/2}$ measured at 1.85 K across the channel at various decay times, where
$\langle{...}\rangle_x$ denotes an ensemble average of the results obtained at location $x$ at each given decay time from the analysis of 100 deformed
tracer line images.
Similar to typical classical grid flows, the quantum flow in the immediate wake of the grid is not perfectly homogeneous and isotropic. The observed
deformation of the tracer line suggests the existence of large scale eddies spanning the entire width of the channel following the towed grid. This is
most likely caused by mechanical imperfections in the construction of the grid and its support. Nevertheless, this inhomogeneity quickly decays, being
virtually completely eliminated within 4 s. In contrast with the mean flow and its marked initial inhomogeneity, the profile of the velocity variance
$\sigma(x)$ is much more homogeneous, even at small decay times.

\begin{figure}
  \centering
  \includegraphics{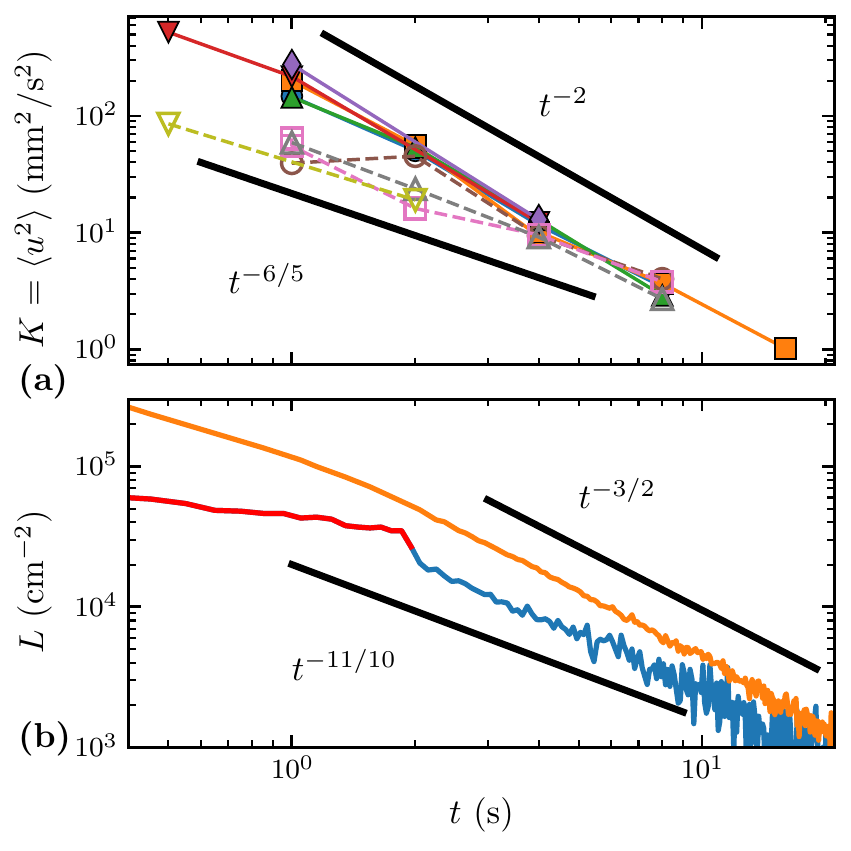}
  \caption{ (a) Decaying turbulent kinetic energy of the normal fluid, $K(t)$, and (b) vortex line density, $L(t)$, originating from towing the grid
    at 50 mm/s - empty symbols/blue line and 300 mm/s - full symbols/orange line. The energy decay is shown for temperatures 1.45~K (\pcircle), 1.65~K
    (\psquare), 1.85~K (\ptriup), 2.00~K (\ptridown), 2.15~K (\pdiamond). The red line corresponds to the early decay of $L(t)$ for times when the
    grid is still moving. The decays are quasi-classical in character. The early part of the decay, when the energy containing length scale
    $\ell_\mathrm{e}$ grows, displays the characteristic decay exponents $K(t)\propto t^{-6/5}$, $L(t)\propto t^{-11/10}$, while the late universal
    part of the decay, when $\ell_\mathrm{e}$ is saturated by the channel size, obeys $K(t)\propto t^{-2}$, $L(t)\propto t^{-3/2}$
    \cite{Stalp1999,Skrbek2000}. These decay rates are illustrated by thick black lines. For towed grid velocity of 300 mm/s saturation occurs too
    early for the early part of the decay to be resolved.  The shown data are obtained at 1.85 K.}
  \label{fig:energy-decay}
\end{figure}

Despite the initial transient inhomogeneity at large scales, the temporal decays of the normal fluid turbulent kinetic energy,
$K(t)=\langle{\sigma^2}\rangle$, and the vortex line density in the superfluid, $L(t)$, exhibit clear decay characteristics of quasi-classical
homogeneous isotropic turbulence. As discussed in detail in Refs. \cite{Stalp1999,Skrbek2000}, in the early decay stage of grid turbulence when the
energy containing length scale $\ell_\mathrm{e}$ grows from the injection scale (i.e., comparable to the mesh size) to the channel width, the
characteristic decay exponents for quasi-classical homogeneous isotropic turbulence should be $K(t)\propto t^{-6/5}$ and $L(t)\propto t^{-11/10}$; in
the late universal decay stage after $\ell_\mathrm{e}$ is saturated by the channel width, $K(t)\propto t^{-2}$ and $L(t)\propto t^{-3/2}$ should be
expected. These decay behaviors are clearly observed in our data. Note that at high towed-grid velocity (i.e., $v_g=300$ mm/s), the saturation of
$\ell_\mathrm{e}$ likely occurs too rapidly for the early decay stage to be resolved. Furthermore, the transient inhomogeneity at small decay times
may also affect the decay characteristics in this regime. At the lower grid velocity (i.e., $v_g=50$ mm/s), the late universal decay stage appears at
relatively large decay times (i.e., over 3--4 s) due to the slower increase of $\ell_\mathrm{e} = \ell_\mathrm{e}(t)$ \cite{Stalp1999,Skrbek2000}.

\subsection{Transverse velocity structure functions}
The observed quasi-classical decay laws for $K(t)$ and $L(t)$ suggest that classical K41-like scalings in other turbulence statistics such as the
velocity structure functions may also be expected. For instance, for fully developed classical homogeneous isotropic turbulence, the second order
transverse velocity structure function, defined as
\begin{equation}
S_2^\bot(r) = \mean{|v_y(x+r) - v_y(x)|^2}\,, 
\label{eq:S2}
\end{equation}
should scale with the transverse separation distance $r$ as $S_2^\bot(r) \propto r^{2/3}$ \cite{Henze-book-1975}.

\begin{figure}
  \centering
  \includegraphics{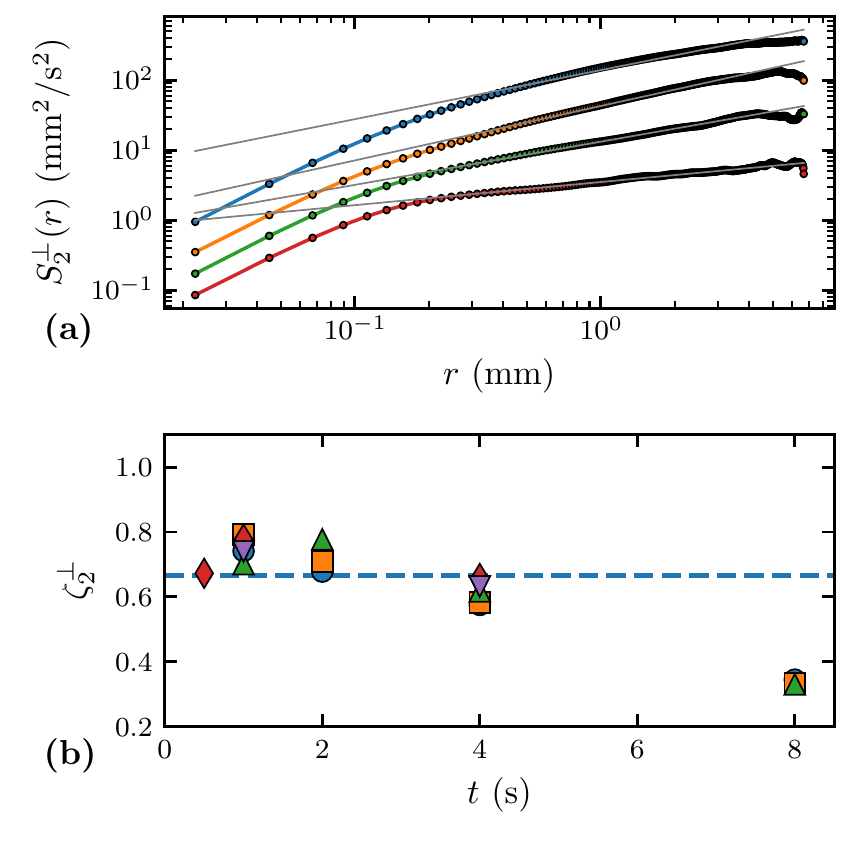}
  \caption{(a) Calculated second order transverse velocity structure functions, $S_2^\bot(r)$, for $T=1.85$~K at decay times (from up to down) $t=$1 s
    ({\color{pblue}\pcircle}); 2 s ({\color{porange}\pcircle}); 4 s ({\color{pgreen}\pcircle}); and 8 s ({\color{pred}\pcircle}).  The grid velocity
    is $v_g=300$ mm/s. The grey solid lines represent power-law fits to the data in the range 0.2 mm $\leq{r}\leq$~4 mm.  (b) The scaling exponent
    $\zeta_2^\bot$ deduced from the power-law fits, such as shown in (a), at different temperatures: 1.45~K ({\color{pblue}\pcircle}), 1.65~K
    ({\color{porange}\psquare}), 1.85~K ({\color{pgreen}\ptriup}), 2.0~K ({\color{pred}\pdiamond}), 2.15~K ({\color{pviolet}\ptridown}). The dashed
    horizontal line shows the K41 scaling $\zeta_2^\bot = 2/3$.}
  \label{fig:S2-decay}
\end{figure}

In the case of He II grid turbulence, the situation is more complex. Fig.~\ref{fig:S2-decay} (a) shows typical examples of calculated $S_2^\bot(r)$
curves, for $T=1.85$~K with a grid velocity of $v_g=300$ mm/s at decay times $t=$1, 2, 4, and 8 s.
Non-trivial power-law scalings of $S_2^\bot(r)$ are clearly observed in the scale range 0.2 mm $\leq{r}\leq$~4 mm. The quadratic-like dependence of
$S_2^\bot(r)$ at small $r$ is probably caused by smearing of the measured velocity field limited by the width of the tracer line (i.e., about 100
$\mu$m) rather than due to the viscous flow. By fitting the data in 0.2 mm $\leq{r}\leq$~4 mm with a power-law form
$S_2^\bot(r)\sim{r^{\zeta_2^\bot}}$, the scaling exponent $\zeta_2^\bot$ can be extracted and is shown in Fig.~\ref{fig:S2-decay} (b). Data at other
temperatures are also included in this figure. We see that the data display slightly steeper than K41 scaling (i.e., $\zeta_2^\bot>2/3$) for the 1 s
and 2 s measurements and shallower than K41 (i.e., $\zeta_2^\bot<2/3$) for 8 s and later measurements. We note in passing that this behavior is not
unusual in classical decaying grid turbulence, especially before the wakes of individual bars of the grid fully coalesce
\cite{ComteBellot1966,Sreenivasan1997}. An additional factor to consider is possible parasitic radiative heating to the channel. This parasitic
heating can cause weak thermal counterflow which may become important at long decay times when the grid turbulence strength is low.

\begin{figure}
  \centering
  \includegraphics[width=0.9\linewidth]{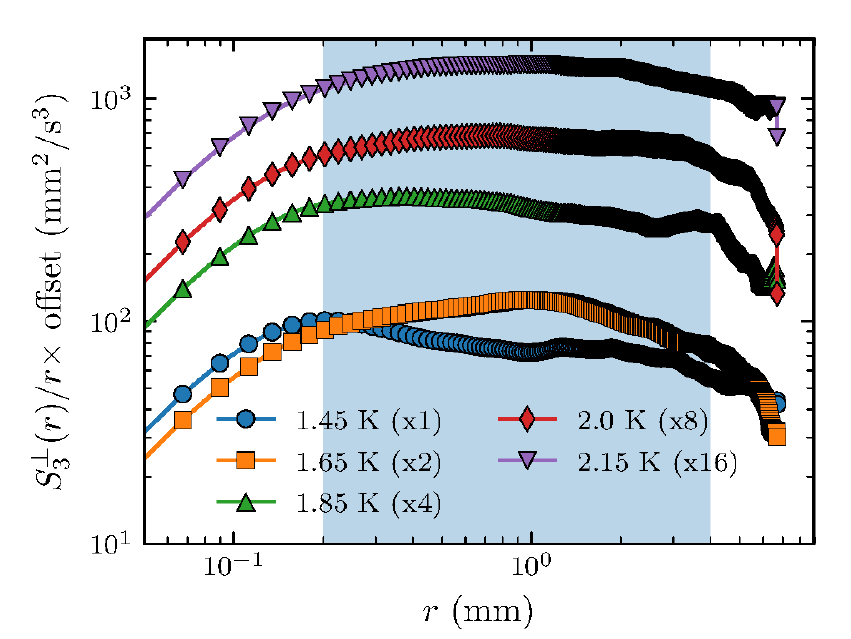}
  \caption{Third order transverse velocity structure functions compensated by linear scaling, $S_3^\bot(r)/r$, plotted versus the separation distance
    $r$. Data for 300 mm/s grid velocity and 4 s decay time are shown for temperatures and offsets as indicated.}
  \label{fig:S3}
\end{figure}

Besides the second order structure function, the Kolmogorov 4/5-law also states that within the inertial range of scales, the third order longitudinal
velocity structure function should be given by
\begin{equation}
S_3^{\parallel}(r) = (-4/5)\varepsilon r \,,
\label{eq:KolmLaw}
\end{equation}
where $\varepsilon=-dK/dt$ is the energy dissipation rate \cite{Frisch1995book,Noullez1997}. In our experiment, only the transverse velocity structure
functions $S_n^\bot$ are accessible. Nevertheless, it can be shown \cite{Frisch1995book} that the scaling is equal for both $S_2^\bot$ and
$S_2^\parallel$ structure functions in three dimensional incompressible homogeneous isotropic turbulence and that the Kolmogorov 4/5-law ought to be
valid also for the transverse structure function \cite{Lvov1997,Chkhetiani1996}. On the other hand, there is an experimental evidence that the scaling
exponent of $S_3^\bot$ in high Reynolds (Re) number atmospheric turbulence is slightly less (perhaps due to finite Re) but very close to unity
\cite{Dhruva1997}. We have evaluated the 3rd order transverse structure function $S_3^\bot(r) = \mean{|v_y(x+r) - v_y(x)|^3}$ at 4 s decay time where
classical scaling is clearly observed for $S_2^\bot(r)$ as shown in Fig.~\ref{fig:S2-decay}. The calculated values of $S_3^\bot(r)/r$ as a function of
$r$ are shown in Fig.~\ref{fig:S3} at various temperatures. Over a similar range, 0.2 mm $\leq{r}\leq$~4 mm, we see a reasonably good linear
dependence of $S_3^\bot(r)$ on $r$, which coincides with the Kolmogorov 4/5 law in the inertial cascade range. Similar behavior is observed at 4 s for
the other available temperatures and for both grid velocities, however, for decay times other than 4 s any linear scaling of $S_3^\bot(r)$ cannot be
convincingly resolved.

\begin{figure}
  \centering
  \includegraphics[width=0.9\linewidth]{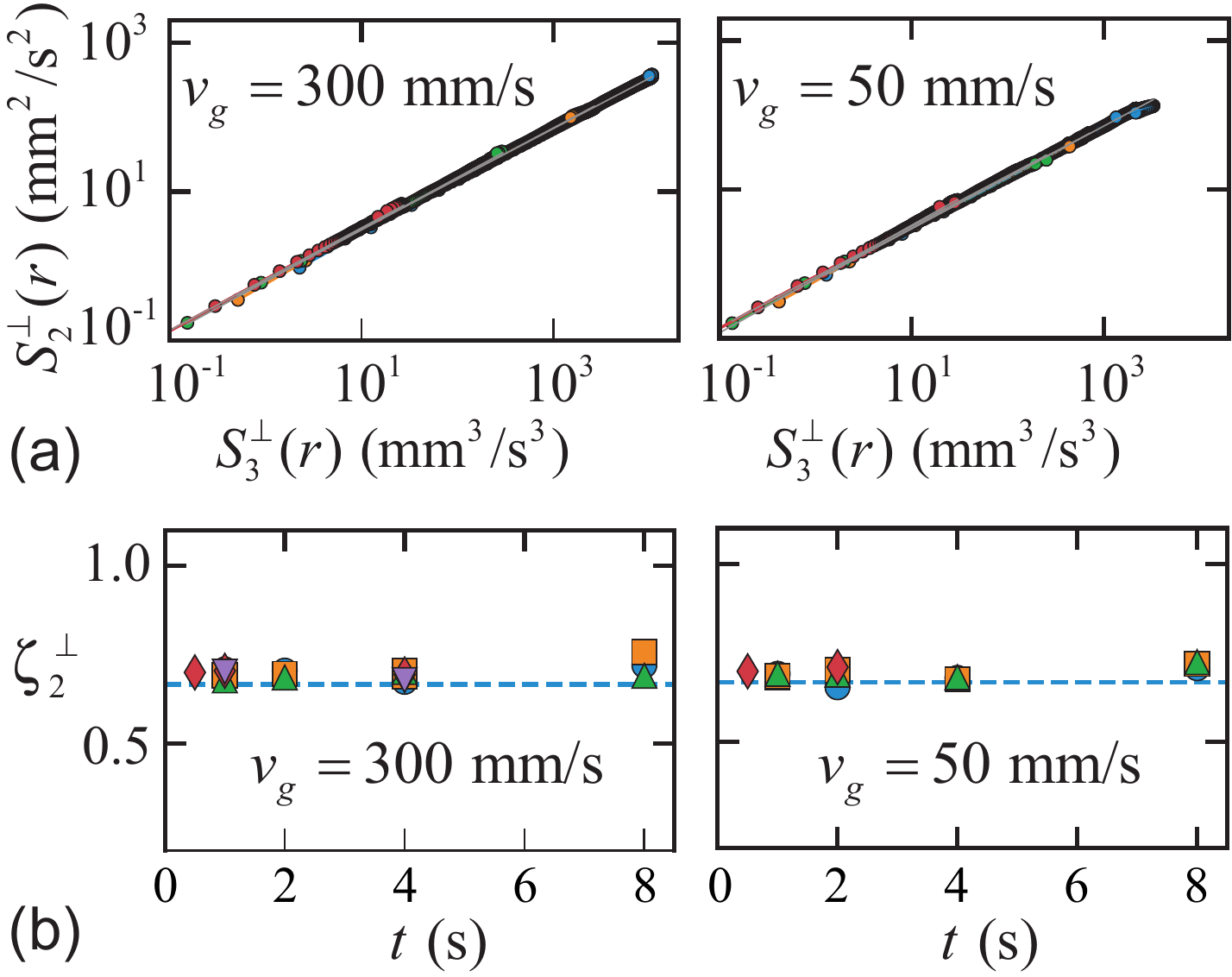}
  \caption{(a) Extended self-similarity scaling of $S_2^\bot(S_3^\bot)$ for data obtained at $T=1.85$ K and decay time $t=4$ s. (b) The scaling
    exponent $\zeta_2^\bot$ extracted using the extended self-similarity hypothesis at various color coded temperatures: 1.45~K
    ({\color{pblue}\pcircle}), 1.65~K ({\color{porange}\psquare}), 1.85~K ({\color{pgreen}\ptriup}), 2.0~K ({\color{pred}\pdiamond}), 2.15~K
    ({\color{pviolet}\ptridown}). The dashed horizontal lines show the K41 scaling $\zeta_2^\bot = 2/3$.  }
  \label{fig:S2-ESS}
\end{figure}

The scaling exponents of the structure functions can also be obtained by using the so-called extended self-similarity hypothesis
\cite{Benzi1991}. This hypothesis states that the scaling of a structure function $S_n(r)$ in the inertial scale range should be equivalent to the
scaling of $S_n(r)\propto\left(S_3(r)\right)^{\zeta_n}$. Indeed, structure function scalings based on extended self-similarity appear to be very
robust and can extend down to the dissipative scale range even for turbulent flows with moderate Reynolds numbers \cite{Benzi-EPL-1993}, therefore
allowing for significant improvement in experimental determination of the scaling exponent $\zeta_n$ \cite{Dubrulle-PRL-1994}. In
Fig.~\ref{fig:S2-ESS} (a), we show $S_2^\bot(r)$ versus $S_3^\bot(r)$ on a log-log plot for the data obtained at 1.85 K at 4 s decay time. For both
grid velocities, a linear dependence of $\log S_2^\bot(r)$ on $\log S_3^\bot(r)$ is clearly seen and extends to a wide range of length scales. The
values of the scaling exponent $\zeta_2^\bot$ deduced using the extended self-similarity hypothesis at various decay times and temperatures are shown
in Fig.~\ref{fig:S2-ESS} (b), which display noticeably improved agreement with the K41 scaling.

\begin{figure}
  \centering
  \includegraphics[width=0.9\linewidth]{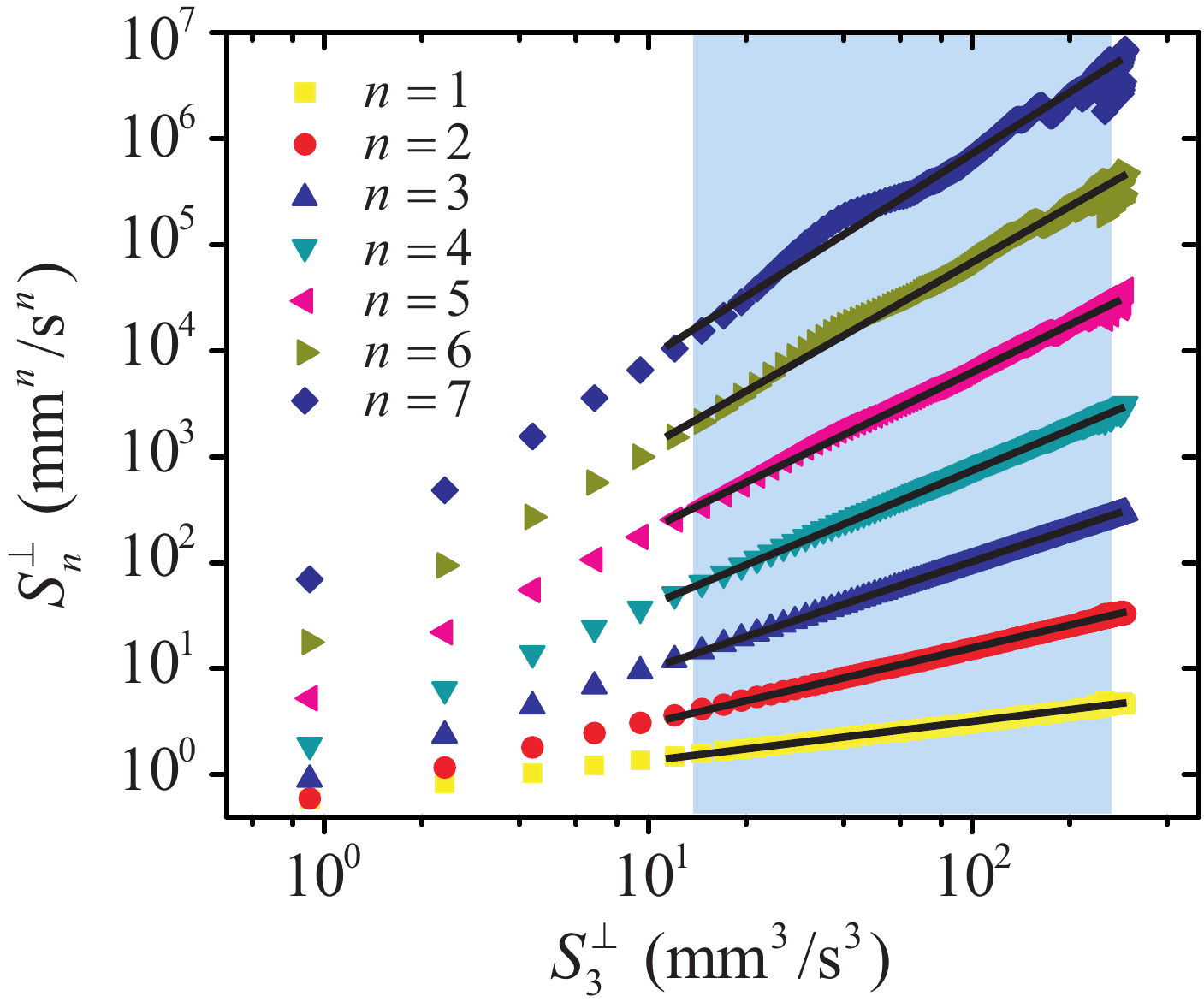}
  \caption{Extended self-similarity. Transverse velocity structure functions $S_n^\bot$ for $n=1-7$ are plotted versus $S_3^\bot$. The black lines are
    linear fits of $\log S_n^\bot$ vs. $\log S_3^\bot$ to the data that fall within the shadowed region that corresponds to the shadowed region in
    Fig.~\ref{fig:S3} where $S_3^\bot/r$ appears to be flat. The particular case shown is for 1.85 K, 300 mm/s grid velocity and 4 s decay time. Other
    cases appear qualitatively similar.  }
  \label{fig:ess}
\end{figure}

\subsection{Temperature dependence of intermittency corrections}
Turbulence intermittency is normally evaluated by statistical analysis of the experimental data via higher order structure functions $S_n(r)$ that are
more sensitive to the occurrence of rare events. The transverse velocity structure function of order $n$ is defined through the transverse velocity
increments as
\begin{eqnarray}
\label{eq:Sn-pdf-definition}
  S_n^\bot(r) =& \mean{|\delta v_y(r)|^n} =& \int_{-\infty}^\infty\dd x |x|^n\pdf_r(x),
\end{eqnarray}
where $\pdf_r(x)$ represents the probability density function of $\delta v_y(r)$. In order for $S_n^\bot$ to be evaluated accurately, the experimental
estimation of the PDF needs to have well-resolved tails because of the $x^n$ term in the integral, which in turn requires very large data sets. Our
setup does not presently allow for the collection of very large data sets. Typical data sets are limited to about $10^4$ samples. Another issue is
that, although the individual He$_2^*$ molecules are of nm size and are true tracers of normal fluid flow, we cannot detect individual tracers - a
large number of them closely spaced are needed to satisfy our sensitivity limit. Rare events resulting in large departures of individual tracers are
therefore invisible to us. In other words, our experimentally resolved length scale is limited by the thickness of the deformed tracer line,
$\ell_{exp}\simeq 100$ $\mu$m. A more detailed discussion of the uncertainties associated with the calculated structure functions is provided in the
Appendix.

According to the K41 theory, for fully developed homogeneous isotropic turbulence in classical fluids without any intermittency, the structure
function in the inertial cascade range should scale as $S_n(r) \propto r^{\zeta_n}$ with the scaling exponent $\zeta_n=n/3$
\cite{Henze-book-1975}. Intermittency in real turbulent flows of conventional viscous fluids leads to corrections of the scaling exponents, and this
correction becomes more pronounced at large $n$. In order to reliably determine the actual scaling exponents of the transverse structure functions
$\zeta_n^\bot$ in our quantum grid turbulence, we again utilize the extended self-similarity hypothesis. Furthermore, we focus our study on data
obtained at 4 s decay time, since the scalings of $S_2^\bot(r)$ and $S_3^\bot(r)$ presented in the previous section suggest fully developed
homogeneous isotropic turbulence at this decay time.

In Fig.~\ref{fig:ess}, the calculated $S_n^\bot(r)$ versus $S_3^\bot(r)$ for $n=1$ to 7 are shown for data obtained at 1.85 K with a grid velocity
$v_g=300$ mm/s. Clear power law dependence of $S_n^\bot(r)$ on $S_3^\bot(r)$ is seen, which extends to the smallest scales probed in the
experiment. Data obtained at other temperatures appear qualitatively similar. We then perform a power-law fit of the form
$S_n^\bot(r)\propto\left(S_3^\bot(r)\right)^{\zeta^\bot_n}$ to the data (shown as black lines in Fig.~\ref{fig:ess}). The fit is restricted to the
range of scales 0.2 mm $<r<$ 4 mm where $S_3^\bot(r)/r$ is reasonably flat, supporting the existence of an inertial cascade.

\begin{figure}
  \centering
  \includegraphics{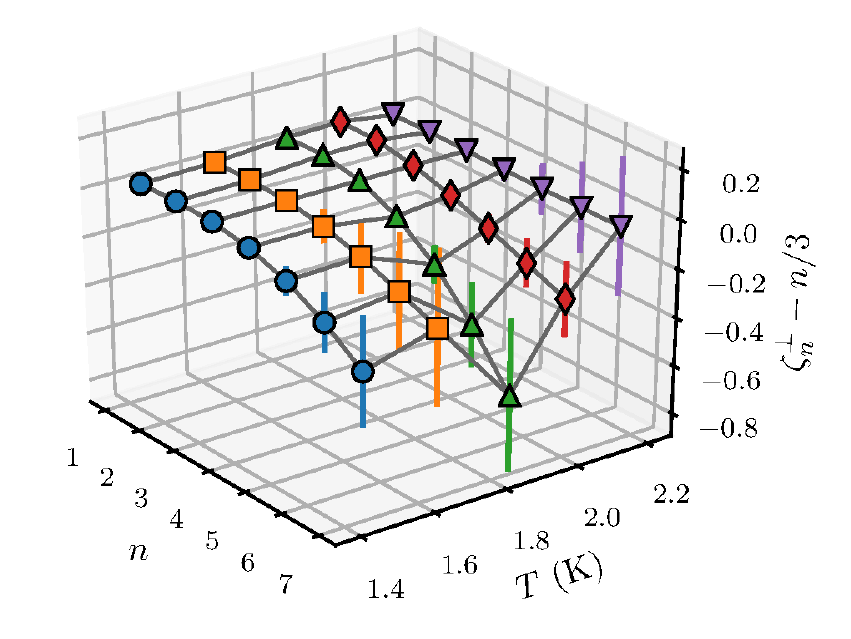}
  \caption{Intermittency corrections to the scaling exponents of the transverse structure functions deduced through extended self-similarity for data
    obtained at 4 s decay time and with grid velocity $v_g=300$ mm/s. The 3D plot shows the temperature dependent deviation of scaling exponents from
    K41 scaling -- 1.45 K ({\color{pblue}\pcircle}), 1.65 K ({\color{porange}\psquare}), 1.85 K ({\color{pgreen}\ptriup}), 2.00 K
    ({\color{pred}\pdiamond}), 2.15 K ({\color{pviolet}\ptridown}).  }
  \label{fig:scaling-exponents}
\end{figure}

The deduced scaling exponents $\zeta_n^\bot$, for all investigated temperatures, as a function of the order $n$ are shown in
Fig.~\ref{fig:scaling-exponents}. This figure represents the central result of our work. It is remarkable that the deduced scaling exponents closely
follow the recent theoretical prediction of Biferale \emph{et al.}~\cite{Biferale2017}, i.e., temperature dependent intermittency corrections of the
structure function scaling exponents with a maximum deviation from the K41 scaling at 1.85 K. It should be noted that, while the result for $t = 4$ s
is robust, for small decay times (for additional discussion see the Appendix) and for slower grid velocity the conclusion is not as clear, which is
likely due to insufficiently developed turbulence.

\section{Discussion}
Let us compare our results with similar experimental data available. The recent Grenoble measurements of Rusaouen \emph{et al.}~\cite{Rusaouen2017} in
the wake of a disk in the two-fluid region of superfluid $^4$He found no appreciable temperature dependence in intermittency corrections. The results
of the Grenoble experiment and our experiment therefore appear to be controversial. Nevertheless, there are several reasons why the two experiments
may show different results. First, the prediction of temperature dependent enhanced intermittency is explained by the authors of
ref. \cite{Boue2013a,Biferale2017} via a flip-flop scenario -- a random energy transfer between the normal and superfluid components due to mutual
friction. While He$_2^*$ molecules in our experiment probe the normal fluid solely, the cantilever anemometer and pressure probes used in the Grenoble
experiment~\cite{Rusaouen2017} may not sense such a flip-flop exchange of energy, as it probes both fluids simultaneously. Furthermore, the sizes of
the probes used in the Grenoble experiment are typically much larger than the quantum length scale $\ell_\mathrm{Q}$. Indeed, recent particle image
velocimetry visualization experiments by La Mantia \emph{et al.} in Prague~\cite{LaMantia2014,Svancara2017}, utilizing solid hydrogen/deuterium
particles a few $\mu$m in size, reveal a crossover from classical to quantum signatures of turbulence as the probed length scale crosses
$\ell_\mathrm{Q}$. As discussed previously, our smallest accessible length scale $\ell_\mathrm{exp}$ - the width of tracer line - is about 100
$\mu$m. At a decay time of 4 s in our experiment, $\ell_\mathrm{Q}\simeq{L^{-1/2}}$ is also about 100 $\mu$m (see Fig.~\ref{fig:energy-decay}). The
quantum length scale $\ell_\mathrm{Q}$ increases at later decay times as the vortex line density $L(t)$ decays. Therefore, our data sets sample the
velocity field near to or below $\ell_\mathrm{Q}$, where one expects the effect of quantized vorticity to become apparent. In the experiments of
Rusaouen \emph{et al.}~\cite{Rusaouen2017}, taking the outer scale of turbulence to be their channel size $\simeq5$ cm, effective kinematic viscosity
$\nu_\mathrm{eff}\simeq 0.1\kappa$ and following the estimations in Babuin \emph{et al.} \cite{Babuin2014a}, the $\kappa$-based large scale Reynolds
number at 1.85 K is roughly $6\times10^4$. This corresponds to $\ell_\mathrm{Q} \approx 7$ $\mu$m. The cantilever probe has a sensing area of
$32\times375$ $\mu$m, which would translate to more than 100 quantized vortices, even if we neglect the likely increase of $L$ in the vicinity of any
obstacles \cite{Hrubcova2018}. The experiment of Rusaouen \emph{et al.}~\cite{Rusaouen2017} therefore naturally measures the same intermittency
corrections as in classical turbulence.

\section{Conclusions}
We have designed and performed an experiment to study quasiclassical turbulence in the wake of a towed grid in He II, using a recently developed
He$^*_2$ molecular tracer-line tagging velocimetry technique and a traditional second sound attenuation method. Our main result is that, despite the
fact that our data sets are not as large as they ideally ought to be, extended self-similarity reveals temperature dependent intermittency corrections
that peak in the vicinity of 1.85~K, in excellent agreement with recent theoretical predictions \cite{Boue2013a,Biferale2017}. The universality of the
intermittency corrections found in many different turbulent flows of classical viscous fluids \cite{Sreenivasan1997} therefore cannot be extended to
quantum turbulence in superfluid $^4$He. It seems that the role of cliffs that are thought to be responsible for rare but intense events resulting in
intermittency corrections in classical turbulence is at least partly played by quantized vortices in He II. In order to observe this ``quantum"
intermittency, similarly as in classical homogeneous isotropic turbulence, where one has to resolve small scales down to the Kolmogorov dissipation
scale, in quantum turbulence one needs to resolve scales below the quantum length scale $\ell_\mathrm{Q}$.

\begin{acknowledgments}
  We thank V. S. L'vov, K. R. Sreenivasan and W. F. Vinen for fruitful discussions. WG acknowledges the support by the National Science Foundation
  under Grant No. DMR-1807291. The experiment was conducted at the National High Magnetic Field Laboratory, which is supported by NSF Grant
  No. DMR-1644779 and the state of Florida. EV and LS thank the Czech Science Foundation for support under GA\v{C}R 17-03572S.
\end{acknowledgments}

\section*{Appendix: Estimation of Structure Function Errors}
\label{sec:errors}

High order structure functions required to estimate the intermittency corrections are sensitive to rare events -- events of low probability which
would contribute to the ``tails'' of the statistical distribution. In samples of limited size, these tails could be under-resolved, what could lead to
an erroneous estimation of the structure functions. We adopt a simple strategy to estimate these errors due to lack of statistics: an estimate of the
PDF is calculated from the measured data, which is then extended beyond the range of experimental data using a fit to a particular choice of a
heavy-tailed statistical distribution. The difference between the value obtained through Eq.~\eqref{eq:Sn-pdf-definition} using either a
non-extrapolated or extrapolated PDF is then used as the estimate of the error caused by under-resolved tails of the statistical distribution.

\begin{figure}
  \centering
  \includegraphics{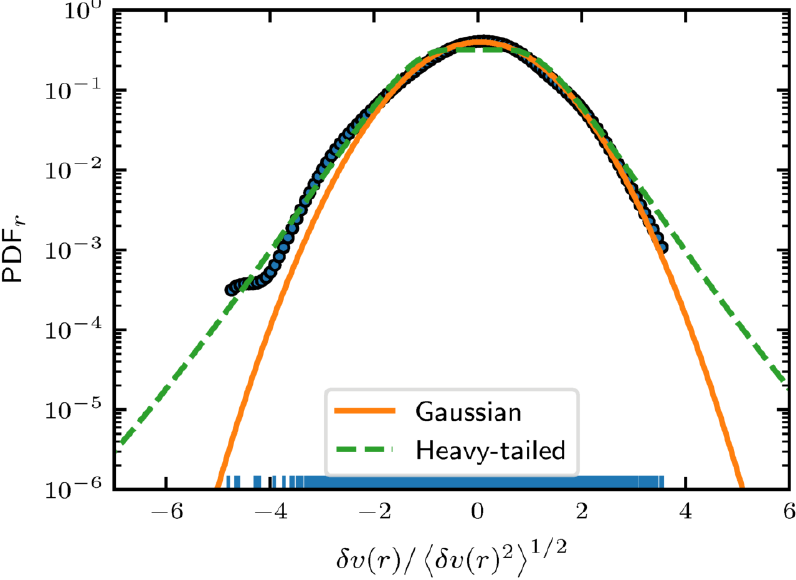}
  \caption{Probability distribution function of the velocity increments. The rug plot shows the actual data set used for the KDE. The data shown is
    for 1.45 K, 300 mm/s grid velocity and 4 s decay time.}
  \label{fig:pdf}
\end{figure}

\begin{figure}
  \centering
  \includegraphics{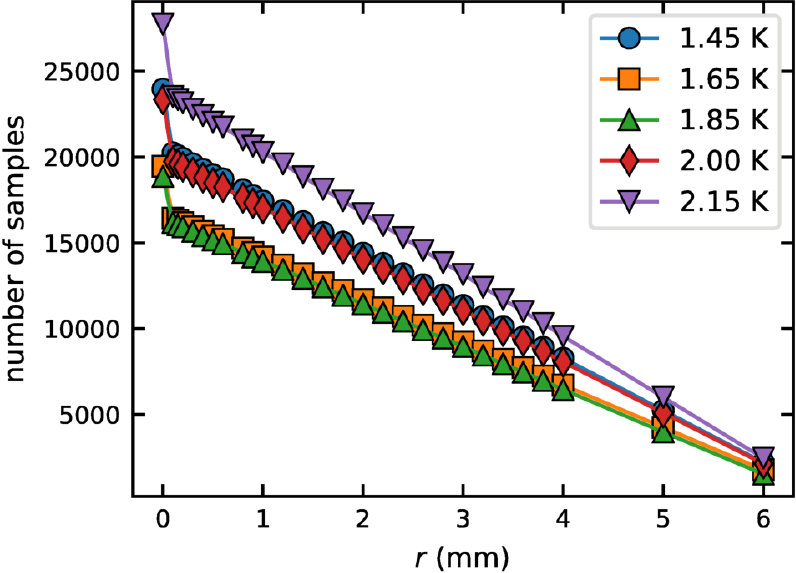}
  \caption{Number of velocity increment samples as a function of separation for 4 s data sets and all experimental temperatures.}
  \label{fig:sample-numbers}
\end{figure}

We calculate an estimation of the PDF from the measured velocity increments using the kernel density estimation (KDE) as
\begin{equation}
  \label{eq:pdf-kde}
  \pdfkde_r(x) = \frac{1}{N}\sum_{i=0}^N \frac{1}{\sqrt{2\pi b }} e^{-(x - \delta v(r)_i)^2/ 2b^2},
\end{equation}
where the sum runs through all $N$ measured samples of $\delta v(r)_i$ at a given separation $r$. The result, for a particular case, is shown in
Fig.~\ref{fig:pdf}. The number of samples for the 4 s decay data sets is in Fig.~\ref{fig:sample-numbers}.

\begin{figure}
  \centering
  \includegraphics{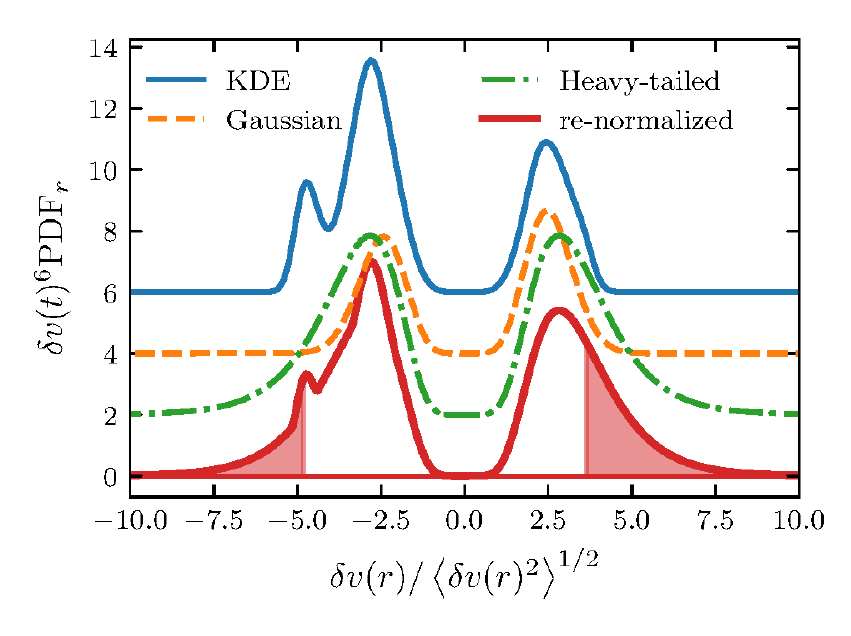}
  \caption{Calculation of the sixth moment of the velocity increment distribution. The curves are offset along the $y$-axis with an offset incrementing by
    2. Same data set as in Fig.~\ref{fig:pdf}.}
  \label{fig:pdf-moment}
\end{figure}

\begin{figure}
  \centering
  \includegraphics{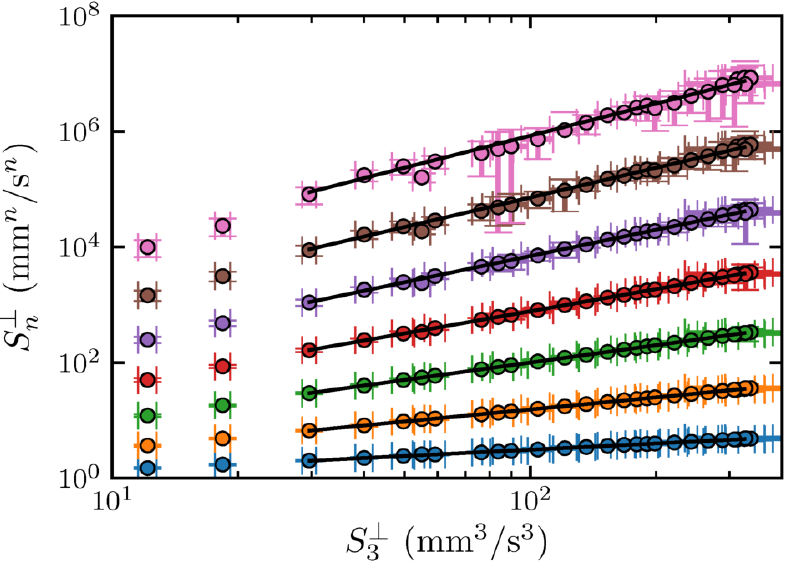}
  \caption{Structure functions $S_n^\bot$ of orders 1 to 7 as a function $S_3^\bot$ at 1.85 K, 4 s decay time and associated error bars calculated
    using the scheme described in Sec.~\ref{sec:errors}. These plots are analogues of curves in Fig.~\ref{fig:ess}.}
  \label{fig:pdf-ess-T185-4s}
\end{figure}

To estimate the error in calculating a given moment, we extrapolate the estimated PDF either by natural extension of the KDE \eqref{eq:pdf-kde}
outside the range of the data set, or by using fits to either the normal (Gaussian) distribution,
\begin{equation}
  \label{eq:normal-dist}
  \pdfn_r(v) = \frac{1}{\sqrt{2\pi s^2}} \exp\left(-\frac{v^2}{2s^2}\right),
\end{equation}
or a particular case of heavy-tailed distribution
\begin{equation}
  \label{eq:heavy-tailed-dist}
  \pdfht_r(v) = \frac{\exp(s^2/2)}{4m}\left[1 - \erf\left(\frac{\log\left(\frac{|v|}{m}\right) + s^2}{\sqrt 2 s}\right)\right],
\end{equation}
where $s$ and $m$ are adjustable parameters. This form of the PDF was found to describe Lagrangian accelerations \cite{Mordant2004a}, but in our case
it is used simply for reasons of convenience (we measure Eulerian transverse velocity increments), as it allows for smooth varying of the weight of
the tails. Note that using a distribution with power law tails would be inconsistent in our case as such a distribution would render the moment of
sufficiently high orders undefinable. Using the two fits and the KDE, we construct a new PDF with the shape of an envelope (point-wise maximum) of the
three estimates. The point-wise maximum breaks the normalization of the probability density function which needs to be re-normalized to the integral
of unity. This effectively decreases the probability in the central peak and moves it towards the tails. An illustration of this procedure is shown in
Fig.~\ref{fig:pdf-moment} for calculating the sixth order moment of a distribution.

As an error estimate of the moment, we take the absolute value of the difference between the moment calculated using the natural extension of the KDE
\eqref{eq:pdf-kde} and the re-normalized \pdf. Graphically, this is given approximately by the area under the tails of the re-normalized \pdf~outside
the range of the data set, shown by the shaded area in Fig.~\ref{fig:pdf-moment}. For calculation of the value of the structure function, we use
\pdfkde. This estimate has a very sharp cutoff (faster than normal distribution) outside the range of the experimental data set (essentially
equivalent to extending a histogram with zeros) so that the value is not affected by any particular choice of extrapolation. The result is shown in
Fig.~\ref{fig:pdf-ess-T185-4s}. We note that the errors of the structure functions render flatness (ratio $S_4^\bot$/$(S_2^\bot)^2$)
unusable for quantitative analysis of intermittency. 

\begin{figure}
  \centering
  \includegraphics{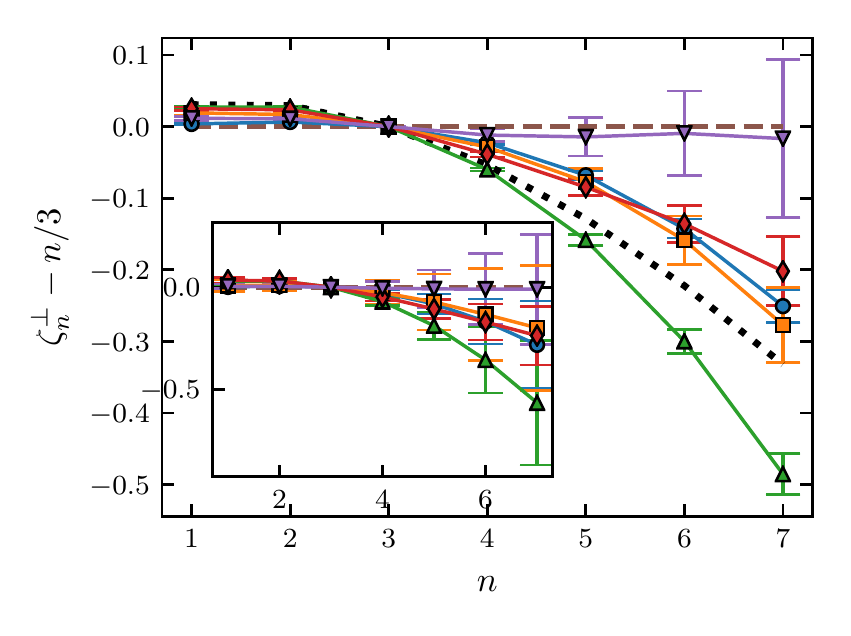}
  \caption{Structure function scaling exponents calculated from PDF-derived structure functions. The error bars are just the standard errors of a
    total least squares linear regression, also known as orthogonal distance regression. The inset shows scaling exponents of structure functions
    calculated using Eq.~\eqref{eq:Sn-pdf-definition}, see the main text for how the error bars are calculated for the data shown in the inset. All
    data for 4 s decay time and 300 mm/s grid velocity. The temperatures are 1.45 K ({\color{pblue}\pcircle}), 1.65 K ({\color{porange}\psquare}),
    1.85 K ({\color{pgreen}\ptriup}), 2.00 K ({\color{pred}\pdiamond}), 2.15 K ({\color{pviolet}\ptridown}). The black dotted line shows the
    prediction of the She-Leveque theory\cite{She-PRL-1994}.}
  \label{fig:zetas-pdf}
\end{figure}
 
We also calculate the structure functions directly from the ensemble average, using the definition Eq. \eqref{eq:Sn-pdf-definition}. The intermittency
corrections resulting from both procedures are shown in Fig.~\ref{fig:zetas-pdf}. Due to the rather arbitrary choice of the heavy-tailed distribution,
the definition of the re-normalized PDF and the definition of the error itself, we also calculate the errors using a bootstrapping scheme
\cite{Efron1981}. The set of all $N$ measured samples entering the calculation of $S_n^\bot$ in Eq.\eqref{eq:Sn-pdf-definition} is sampled at random
(with possible repetitions and omissions) to form $B=5000$ new synthetic sets of length $N$. The standard deviation of the moment
\eqref{eq:Sn-pdf-definition} calculated for these new $B$ data sets is used as the error. The resulting error bars were significantly smaller than
those calculated using the re-normalized PDF and the results were consistent with the straightforward calculation by directly averaging the sample and
are not shown here.

One might justifiably become alarmed by the correlation between the number of samples in Fig.~\ref{fig:sample-numbers} and the deviation from K41
scaling in Fig.~\ref{fig:zetas-pdf}. This, however, appears to be a coincidence. The correlation is not present for other data sets, and artificially
restricting the data sets at 4 s to a random choice (with replacement) of 10000, 5000 or 2000 samples does not have a strong effect on the observed
scaling exponents (although the quality of the structure functions does decrease, as is to be expected). In particular, the minimum near 1.85~K
persists unaffected.


%

\end{document}